# High-temperature growth of GaN nanowires by molecular beam epitaxy: toward the materials quality of bulk GaN

J. K. Zettler,[a] C. Hauswald, P. Corfdir, M. Musolino, L. Geelhaar, H. Riechert, O. Brandt, and S. Fernández-Garrido

*Paul-Drude-Institut für Festkörperelektronik, Hausvogteiplatz 5–7, 10117 Berlin, Germany*

In molecular beam epitaxy, the spontaneous formation of GaN nanowires on Si(111) substrates at elevated temperatures is limited by the long incubation time that precedes nanowire nucleation. In this work, we present three unconventional growth approaches to minimize the incubation time and thus facilitate significantly higher growth temperatures (up to 875°C). We achieve this by: (i) using III/V flux ratios larger than one to compensate for Ga desorption, (ii) introducing a two-step growth procedure, and (iii) using an AlN buffer layer to favor GaN nucleation. The GaN nanowire ensembles grown at so far unexplored substrate temperatures exhibit excitonic transitions with sub-meV linewidths and the low-temperature photoluminescence spectra are comparable to those of state-of-the-art free-standing GaN layers grown by hydride vapor phase epitaxy.

In crystal growth, the concentration of points defects as well as the incorporation of impurities decrease with increasing growth temperature. The optimal growth temperature of III-V compound semiconductors has been theoretically predicted to be close to half of their melting point.[1,2] In plasma-assisted molecular beam epitaxy (PA-MBE), however, the maximum achievable temperature for the growth of GaN films is limited by thermal decomposition because this material is thermodynamically unstable at pressures typical in the molecular beam regime ($< 10^{-4}$ Torr).[3,4] Furthermore, to promote step-flow growth and obtain smooth GaN films, the maintenance of a surfactant Ga-adlayer on the surface is required.[5-7] Due to the exponential increase of Ga desorption with substrate temperature, this requirement implies an additional temperature limitation. For these reasons, the typical substrate temperatures reported for the growth of GaN films by PA-MBE are around 700°C. This value is well below half of the melting point temperature and approximately 300°C lower than the optimal temperatures used in other epitaxial growth techniques, such as metalorganic chemical vapor deposition (MOCVD) or hydride vapor phase epitaxy (HVPE).

Unlike GaN films, the spontaneous formation of GaN nanowires (NWs) in PA-MBE does not require the presence of a Ga-adlayer because these nanostructures form only under N-excess.[8,9] In addition, the N-rich environment allows for a significant reduction in the effective GaN decomposition rate. In fact, GaN decomposition has been found to be negligible during NW elongation up to at least 835°C. Therefore, the growth of GaN in the form of NWs instead of films does not only allow for the growth of single-crystalline GaN on dissimilar substrates but also for the use of significantly higher ($> 100$°C) substrate temperatures.

In this context, it is worth noting that recent experiments demonstrated that the maximum achievable substrate temperature for the growth of GaN NWs on Si(111) substrates is not limited by the elongation stage but by the long delay time that precedes the spontaneous formation of GaN NWs in PA-MBE.[10] This delay time, known as incubation time, strongly depends on the impinging fluxes as well as on the substrate temperature.[10] Thus, the key to achieve even higher growth temperatures consists in reducing the incubation time as much as possible.

In this work, we introduce three different growth approaches conceived to decrease the incubation time and thereby enable the growth of GaN NWs on Si substrates at so far unexplored substrate temperatures (up to 875°C). The first approach consist in leaving the commonly used growth regime of nominally N-rich growth conditions, i. e., a ratio between the impinging Ga and N fluxes lower than one. Instead, we use nominally Ga-rich growth conditions (i. e., in terms of the impinging fluxes) to compensate for the high desorption rate of Ga adatoms at elevated temperatures. In the second approach, we use a two-step growth procedure. There, a lower substrate temperature is used during the first step to favor NW nucleation. Finally, in the third approach, we enhance GaN nucleation by introducing an AlN buffer layer. Independent of the growth approach, we found that the low-temperature photoluminescence (PL) spectra of these high-temperature-grown GaN NWs is comparable to those of state-of-the-art free-standing GaN (FS-GaN) layers grown by HVPE.

All samples were grown on $2^{\text{ff}}$ Si(111) substrates in a MBE system equipped with a radio-frequency $N_2$ plasma source for active N and solid-source effusion cells for Ga and Al. The impinging Ga and N fluxes were calibrated in equivalent growth rate units of nm/min, as described in detail in Ref. 6. A growth rate of 1 nm/min is equivalent to $7.3 \times 10^{13}$ atoms/cm$^2$. The substrate temperature was measured using an optical pyrometer calibrated with the $1 \times 1$ to $7 \times 7$ surface reconstruction transition temperature of Si(111) ($\approx 860°$).[11] The as-received Si substrates were etched using diluted (5%) HF. Prior to growth, the substrates were first outgassed at 885°C for 30 min to remove any residual $Si_xO_y$ from the surface. Afterwards, the substrates were exposed to an active nitrogen flux of $\Phi_N = (11.0 \pm 0.5)$ nm/min at the growth temperature for 10 min, except for those samples prepared on AlN-buffered Si. Details about the preparation of AlN buffer layers can be found in Ref. 12. All samples were grown with $\Phi_N = (11.0 \pm 0.5)$ nm/min. Continuous-wave $\mu$-PL experiments were carried out using a He-Cd laser ($\lambda$ = 325 nm) with a power areal density of less than 150 nW/$\mu m^2$. The luminescence was dispersed by an 80 cm Horiba Jobin Yvon monochromator and detected by a charge-coupled device detector. The spectral resolution of the system is $\approx$ 250 $\mu eV$. More details on the PL measurements can be found elsewhere.[13]

---
[a]Electronic mail: zettler@pdi-berlin.de

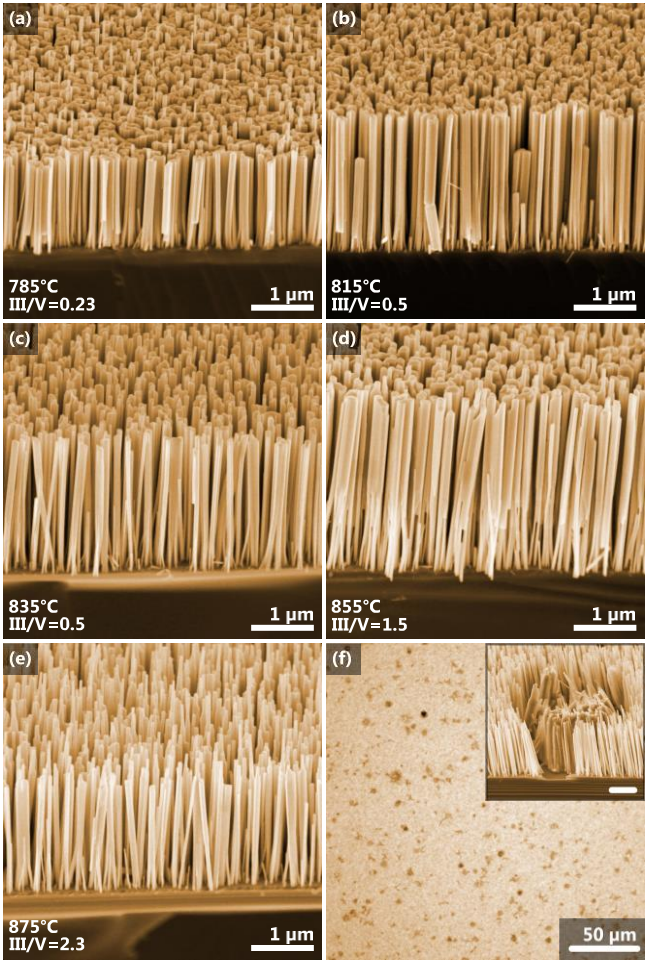

FIG. 1. (a)-(e) Bird-eye scanning electron micrographs of GaN NW ensembles grown on bare Si using different temperatures and III/V flux ratios. The NW ensembles shown in (a)-(c) were grown using N-rich growth conditions while those shown in (d) and (e) were grown using nominally Ga-rich growth conditions. (f) Plan-view optical microscope image of the sample grown at 875°C with III/V= 2.3. The inset shows a bird-eye scanning electron micrographs taken on one of the dark spots observed at the optical microscope. The scale bar of the inset corresponds to 1$\mu m$.

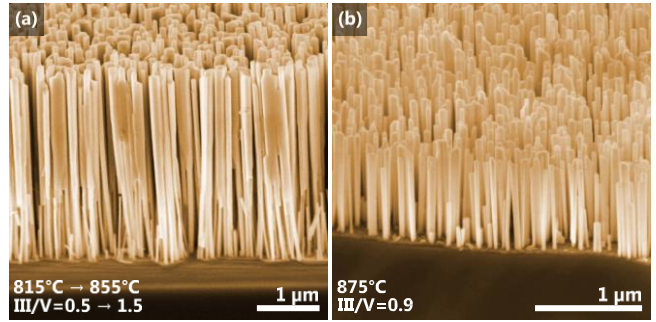

FIG. 2. Bird-eye scanning electron micrographs of GaN NW ensembles prepared on Si at elevated substrates. In (a) we use a two-step growth scheme and in (b) we introduced an AlN buffer layer nucleation to enhance GaN nucleation.

Figures 1(a)-(c) show scanning electron micrographs of GaN NW ensembles grown on bare Si at substrate temperatures between 785°C and 835°C. These three samples were grown using the conventional growth approach, namely, a constant III/V flux ratio lower than one. The growth time for the samples grown at 785 and 815°C was $\approx$ 4 hours. In contrast, for the sample grown at 835°C, we needed to increase the growth time up to 7.5 hours to obtain NWs with an average length comparable to that of the sample grown at 815°C. As discussed in Refs., the extra time was needed to compensate for the exponential increase in the incubation time caused by the higher substrate temperature. Consequently, for the III/V flux ratio used in these two samples (0.5), growth at even higher temperatures becomes unfeasible.

Figures 1(d) and (e) illustrate the results obtained when the substrate temperature is increased by using the first growth approach proposed in this work, i. e., the use of nominally Ga-rich growth conditions. As can be seen in the figures, this growth approach allow us to grow GaN NWs at substrate temperatures as high as 875°C in reasonable times (the growth times for the samples grown at 855 and 875°C were 7 and 8 hours, respectively). As obvious from the micrographs shown in Fig.1, these GaN NWs ensembles are of similar density and morphology as those grown under N-rich growth conditions at lower temperatures. We stress that, despite of the large III/V flux ratios (up to 2.3) used to induce the formation of GaN NWs at 855 and 875°C, the growth still takes place under N-excess because of the high desorption rate of Ga adatoms. Otherwise, as explained in Ref., a compact layer would have formed due to NW radial growth. Interestingly, at growth temperatures above 850°C, we observed that Ga reacts with the Si substrate. This phenomena, known as "melt-back etching", is a well-known problem for the growth of GaN on Si at high-temperatures by other epitaxial growth techniques.[14] As shown in Fig.1(f), the effects caused by Ga-induced melt-back etching can be seen in both optical and scanning electron micrographs. At the optical microscope we see a high density of dark spots that, as shown in the scanning electron micrograph, correspond to the formation of holes in the Si substrate. These holes can be as deep as a few hundreds of nanometers and their size as well as density increase with the substrate temperature and the impinging Ga flux. As can also been seen in the inset of Fig.1(f), the holes leads to the tilting and clustering of the adjacent GaN NWs.

For the growth approach introduced above, the maximum achievable substrate temperature is still limited by the long incubation time that precedes the spontaneous formation of GaN NWs. Our second growth approach aims at reducing the incubation time by using a two-step growth scheme. In the first step, conventional growth conditions, i. e., a moderated substrate temperature and a III/V flux ratio lower than one, are used to keep short the incubation time. Once the formation of the first GaN NWs is detected by reflection high-energy electron diffraction (RHEED),[15,16] the growth temperature is increased for the second step of the growth. For this final stage, the III/V flux ratio can be kept constant. However, the use of nominally Ga-rich growth conditions at this stage allows for the use of even higher substrate temperatures. Figure 2(a) presents a scanning electron micrograph of a GaN NW ensemble prepared using the two-step growth scheme. For the first step, we used the conditions employed for the growth of the sample shown in Fig. 1(b). After observing the onset of NW formation by RHEED, the substrate temperature and the III/V flux ratio were increased up to 855˚C and 1.5, respectively. The total growth time for this sample was ≈ 6 h, i. e., 1 h less than for the sample shown in Fig. 1(b). Nevertheless, despite the shorter growth time, both the morphological properties (average length and diameter, number density and coalescence degree) and the low-temperature PL spectra (not shown here) of these two samples are basically identical.

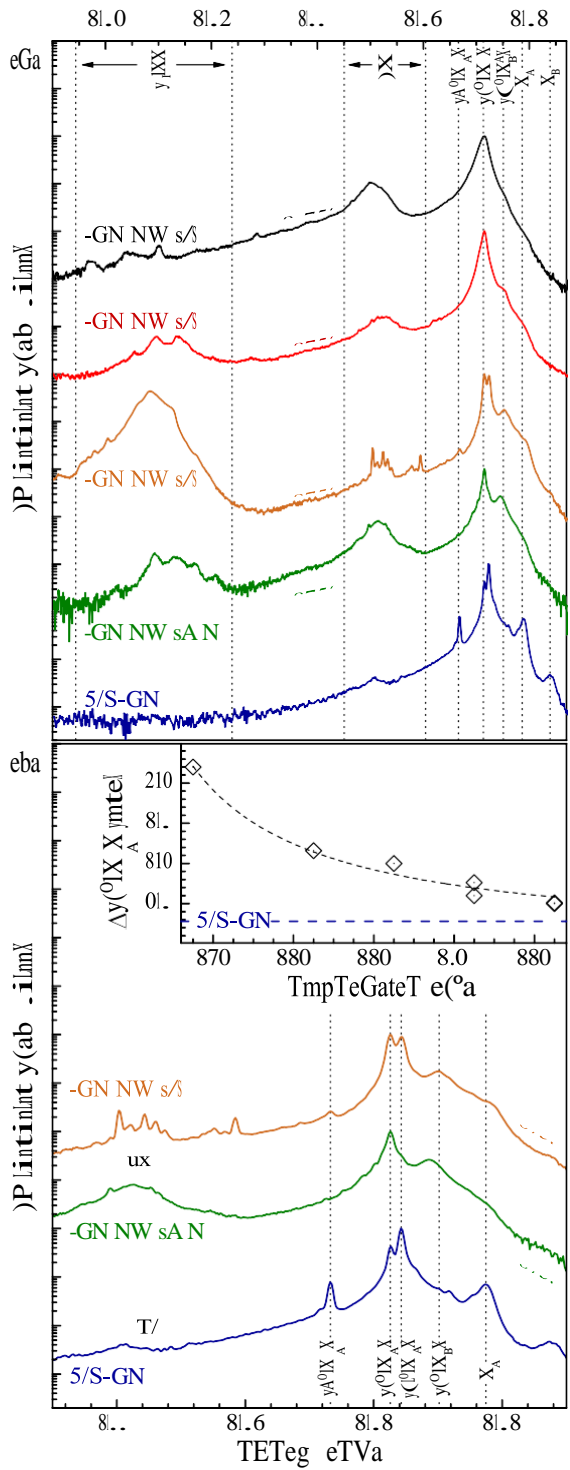

FIG. 3. (a) Normalized and vertically shifted low-temperature (10 K) PL spectra of GaN NW ensembles grown on bare and AlN-buffered Si substrates at different temperatures. (b) Low-temperature PL spectra of the near band-edge region of GaN NW ensembles grown at 875˚C on bare and AlN-buffered Si. The inset shows the variation in the linewidth of the $(O^0,X_A)$ transition with the growth temperature for the GaN NW ensembles prepared on bare Si. For comparison purposes, the PL spectrum of a state-of-the-art FS-GaN layer is also included in (a) and (b). The horizontal dashed line in the inset of Fig. (b) indicates the linewidth of the $(O^0,X_A)$ for the FS-GaN layer.

Our third growth approach exploits the fact that the incubation time does not only depend on the growth conditions but also on the type of substrate.[12,17–19] For instance, it has been shown before that the introduction of an AlN buffer layer favors GaN nucleation on Si substrates.[12,17,18] Therefore, for a single-step growth approach and identical impinging fluxes, one would expect to achieve higher growth temperatures on AlN-buffered Si substrates. Figure 2(b) demonstrates that this is actually the case. As can be seen in the figure, the introduction of a 26 nm thick AlN buffer layer enables the direct growth of GaN NWs at temperatures as high as 875˚C under N-rich growth conditions (III/V = 0.9). Of course, combining this approach with the use of nominally Ga-rich conditions facilitates even higher growth temperatures.

The results presented so far show that, by using non-conventional growth approaches, it is possible to grow GaN NWs at significantly higher temperatures than those previously reported in the literature. The high temperature growth of GaN NWs is, however, accompanied by Ga-induced melt-back etching of the Si substrate. This phenomenon locally influences the morphology as well as the orientation and distribution of GaN NWs. The melt-back etching is more pronounced in those samples grown under nominally metal-rich growth conditions but still present in all other NW ensembles prepared at substrate temperatures above $XXX$˚C. In the following, we examine the influence of the growth temperature on the optical properties of GaN NW ensembles by low-temperature (10 K) PL spectroscopy.

Figure 3(a) shows the near band-edge PL spectra of GaN NW ensembles grown at different temperatures on either bare or AlN-buffered Si substrates. For comparison purposes, we also included the PL spectrum of a state-of-the-art 500 $\mu$m thick free-standing GaN (FS-GaN) layer grown by HVPE. The dislocation density of the FS-GaN layer provided by SINANO is of the order of $XX \times 10^x$ cm². As can be seen in the figure, independent of the substrate temperature, the PL spectra of the NW ensembles are dominated by the donor-bound exciton transition ($D^0, X_A$) at 3.471 eV. For the samples grown at temperatures up to 835℃, this transition is accompanied by a feature-less exponential tail toward lower energies. Besides the ($D^0, X_A$) transition, we also observe in all samples the unknown exciton (UX) transition [at the same energy as the two-electron satellite (TES) peak in the FS-GaN] as well as the ($I_1, X$) emission from stacking faults (SFs). For the samples grown on bare Si, we found that the SF-related luminescence monotonically increases with increasing substrate temperature.

Figure 3(b) shows, in more detail, the PL spectra around the ($D^0, X_A$) transition of the GaN NWs prepared at 875℃ as well as the one of the FS-GaN layer. For all samples, the ($D^0, X_A$) transition is so narrow that we can distinguish, without deconvoluting this line, the individual contribution of neutral O and Si donors, i. e., the ($O^0, X_A$) and ($Si^0, X_A$) transitions. Apart from these lines, we also observe the donor-bound exciton ($D^0, X_B$), the free-exciton $X_A$ and the acceptor-bound exciton ($A^0, X_A$) transitions at higher energies. All these lines can be easily resolved in these NW ensembles because the PL transitions become narrower with increasing substrate temperature. For instance, as can be seen in the inset of Fig. 3(b), the linewidth of the ($O^0, X_A$) transition for the NWs prepared on bare Si monotonically decreases from 2.2 to 0.5 meV when the substrate temperature is increased from 785 to 875℃. The linewidth at 875℃ is thus comparable with the 0.3 meV measured in the FS-GaN layer and much lower than the values reported in the literature for NW ensembles grown at lower temperatures ($1 - 3$ meV).

The narrowing of the PL lines with increasing substrate temperature is accompanied by a change in the intensity ratio between the ($Si^0, X_A$) and ($O^0, X_A$) transitions [see Fig. 3(a)]. Namely, the higher the substrate temperature, the larger the ratio. This fact indicates that Si incorporation is thermally enhanced. As none of these samples was intentionally Si-doped, we attribute Si incorporation to the melt-back etching of the Si substrate. Since the presence of Si adatoms decreases the formation energy of basal plane SFs, the melt-back etching may also explain the monotonic increase in the intensity of the ($I_1, X$) transition with the substrate temperature.

The linewidth of the excitonic transitions in semiconductor crystals scales with the materials quality. In GaN layers, the linewidth of these transitions is determined by the presence of point as well as structural defects. However, in GaN NWs, there is an additional broadening mechanism, namely, the energy dispersion of donor bound-exciton states resulting from their varying distances to the NW sidewall surfaces.[22,23] Interestingly, it has been recently proposed that n-type doping levels on the order of $10^{16}$–$10^{18}$ cm⁻² may lead to the ionization of surface donors and thereby diminished surface induced broadening effects.[13] We believe, that this may be actually the case for our high-temperature GaN NW ensembles which exhibit unprecedentedly narrow donor bound-exciton transitions as well as unintentional doping due the to the melt-back etching of the Si substrate.

If the contribution of surface donors to the photoluminescence of our high-temperature GaN NW ensembles is negligible, the PL-spectra must resemble the materials quality of the NW cores. When comparing the PL spectra of the NW ensemble grown on bare Si at 875℃ with that of the FS-GaN layer (see Fig. 3), we observe that both the position and linewidths of the bound- and free-exciton transitions are quite comparable. Therefore, despite the large lattice-mismatch between Si and GaN (16.9%), these $\approx 2 \mu$m long GaN NWs are free of homogeneous strain and of similar optical quality as 500 $\mu$m thick FS-GaN layers. In addition, due to the better light extraction efficiency for the NW morphology, we found that the integrated PL intensity of the NW ensemble is one order of magnitude higher. Thus, the fabrication of GaN NWs at until now unexplored substrate temperatures along with the likely ionization of surface donors caused by unintentional Si doping have resulted in NW ensembles with unprecedented optical properties. Nevertheless, further studies are needed to disentangle the influences of the substrate temperature and unintentional doping on the PL spectra of high-temperature GaN NW ensembles.

We would like to thank Anne-Kathrin Bluhm for providing the scanning electron micrographs presented in this work, Hans-Peter Schönherr for his dedicated maintenance of the MBE system, and XXX for a critical reading of the manuscript. Financial support of this work by the Deutsche Forschungsgemeinschaft within SFB 951 is gratefully acknowledged.

[1]W. K. Burton, N. Cabrera, and F. C. Frank, Philos. Trans. R. Soc. Lond. A. **243**, 299 (1951).

[2]A. Ishizaka and Y. Murata, J. Phys.: Condens. Matter **6**, L693 (1994).

[3]N. Newman, J. Cryst. Growth **178**, 102 (1997).

[4]S. Fernández-Garrido, G. Koblmüller, E. Calleja, and J. S. Speck, J. Appl. Phys. **104**, 33541 (2008).

[5]T. Zywietz, J. Neugebauer, and M. Scheffler, Appl. Phys. Lett. **73**, 487 (1998).

[6]B. Heying, R. Averbeck, L. F. Chen, E. Haus, H. Riechert, and J. S. Speck, J. Appl. Phys. **88**, 1855 (2000).

[7]G. Koblmüller, S. Fernández-Garrido, E. Calleja, and J. S. Speck, Appl. Phys. Lett. **91**, 161904 (2007).

[8]M. A. Sanchez-García, E. Calleja, E. Monroy, F. Sanchez, F. Calle, E. Muñoz, and R. Beresford, J. Cryst. Growth **183**, 23 (1998).

[9]R. Calarco, R. J. Meijers, R. K. Debnath, T. Stoica, E. Sutter, and H. Lüth, Nano Lett. **7**, 2248 (2007).

[10]S. Fernández-Garrido, J. K. Zettler, L. Geelhaar, and O. Brandt, XXX **XXX**, XXXXX (2014).

[11]T. Suzuki and Y. Hirabayashi, Japanese Journal of Applied Physics **32**, L610 (1993).

[12]M. Musolino, A. Tahraoui, S. Fernández-Garrido, O. Brandt, A. Trampert, L. Geelhaar, and H. Riechert, Nanotechnology **XXX**, XXXXX (2014).

[13]P. Corfdir, J. K. Zettler, C. Hauswald, S. Fernández-Garrido, O. Brandt, and P. Lefebvre, Physical Review B **XXX**, XXXXX (2014).

[14]D. Zhu, D. J. Wallis, and C. J. Humphreys, Reports on Progress in Physics **76**, 106501 (2013).

[15]V. Consonni, M. Knelangen, L. Geelhaar, A. Trampert, and H. Riechert, Physical Review B **81**, 085310 (2010).

[16]V. Consonni, M. Hanke, M. Knelangen, L. Geelhaar, A. Trampert, and H. Riechert, Physical Review B **83**, 035310 (2011).

[17]K. A. Bertness, A. W. Sanders, D. M. Rourke, T. E. Harvey, A. Roshko, J. B. Schlager, and N. A. Sanford, Adv. Funct. Mater. **20**, 2911 (2010).

[18]T. Schumann, T. Gotschke, F. Limbach, T. Stoica, and R. Calarco, Nanotechnology **22**, 095603 (2011).

[19]M. Sobanska, K. Klosek, J. Borysiuk, S. Kret, G. Tchutchulasvili, S. Gieraltowska, and Z. R. Zytkiewicz, J. Appl. Phys. **115**, 043517 (2014).

[20]J. A. Chisholm and P. D. Bristowe, Appl. Phys. Lett. **77**, 534 (2000).

[21]P. Corfdir, C. Hauswald, J. K. Zettler, T. Flissikowski, J. Laḧnemann, S. Ferna´ndez-Garrido, H. T. Grahn, and O. Brandt, Physical Review B **XXX**, XXXXX (2014).

[22]P. Corfdir, P. Lefebvre, J. Ristic´, P. Valvin, E. Calleja, A. Trampert, J.-D. Ganie`re, and B. Deveaud-Ple´dran, J. Appl. Phys. **105**, 013113 (2009).

[23]O. Brandt, C. Pfu¨ller, C. Che`ze, L. Geelhaar, and H. Riechert, Physical Review B **81**, 45302 (2010).